\documentclass{article}

\usepackage{amsfonts}
\usepackage{amssymb}
\usepackage{amsthm}
\usepackage[fleqn]{amsmath}
\usepackage[mathcal]{euscript}
\usepackage{mathrsfs}

\usepackage[latin1,ansinew]{inputenc}
\usepackage{fullpage}
\usepackage{psfrag}
\usepackage{setspace}

\usepackage{graphicx}

\newtheorem{thm}{Theorem}[section]

\newtheorem{rem}[thm]{Remark}

\newcommand{\be}{{\boldsymbol{e}}}

\newcommand{\bg}{{\boldsymbol{g}}}

\newcommand{\bA}{{\boldsymbol{A}}}

\newcommand{\bB}{{\boldsymbol{B}}}
\newcommand{\bE}{{\boldsymbol{E}}}
\newcommand{\bF}{{\boldsymbol{F}}}

\newcommand{\bM}{{\boldsymbol{M}}}

\newcommand{\cM}{{\mathcal{M}}}
\newcommand{\cN}{{\mathcal{N}}}
\newcommand{\cS}{{\mathcal S}}

\newcommand{\Cset}{{\mathbb{C}}}
\newcommand{\Nset}{{\mathbb{N}}}
\newcommand{\Rset}{{\mathbb{R}}}

\newcommand{\Zset}{{\mathbb{Z}}}

\newcommand{\diag}{{\mathrm{diag}}}

\newcommand{\fO}{\mathfrak{O}}

\newcommand{\fD}{\mathfrak{D}}
\newcommand{\fR}{\mathfrak{R}}

\newcommand{\fl}{\mathfrak{l}}
\newcommand{\fm}{\mathfrak{m}}

\begin{document}

\title{Dirac's Point Electron in the zero-Gravity Kerr--Newman World} %

\author{Michael K.-H. Kiessling and A. Shadi Tahvildar-Zadeh\\
  Department of Mathematics\\
  Rutgers, The State University of New Jersey\\
  Piscataway, NJ, USA\\
{miki@math.rutgers.edu, shadi@math.rutgers.edu}
 }

\thanks{This is an expanded version of the talk titled ``The Dirac equation and the Kerr--Newman spacetime,''
given by the first author at the Quantum Mathematical Physics conference, Regensburg, 2014. }

\date{Version of May 10, 2015 (revised version)}
\maketitle


\begin{abstract}
       The results of a study of Dirac's Hamiltonian for a point electron in the zero-gravity Kerr--Newman spacetime 
are reported; here, ``zero-gravity'' means $G\to 0$, where $G$ is Newton's constant of universal gravitation, and
the limit is effected in the Boyer--Lindquist coordinate chart of the maximal analytically extended,
topologically nontrivial, Kerr--Newman spacetime.
       In a nutshell, the results are: the essential self-adjointness of the Dirac Hamiltonian; the reflection symmetry about 
zero of its spectrum; the location of the essential spectrum, exhibiting a gap about zero; and (under two smallness 
assumptions on some parameters) the existence of a point spectrum in this gap, corresponding to bound states of Dirac's
point electron in the electromagnetic field of the zero-$G$ Kerr--Newman ring singularity.
       The symmetry result of the spectrum extends to Dirac's Hamiltonian for a point electron in a generalization 
of the zero-$G$ Kerr--Newman spacetime with different ratio of electric-monopole to magnetic-dipole moment.
       The results are discussed in the context of the general-relativistic Hydrogen problem. 
       Also, some interesting projects for further inquiry are listed in the last section.
\end{abstract}

\section{Introduction}

 There are many studies of Dirac's wave equation on curved background spacetimes, see e.g.
\cite{Erwin32,McVittie,BriCoh,Chandra76a,Chandra76b,Page76,Bel98,CohPow,Zecca,BelMar99,BelCac2010,WinYamA,WinYamB,BSW,FKSYinCPAM2000a,FKSYinCPAM2000b,FKSYinCMP2002,FKSYinATMP2003,Mel00,CacDor}.
 The papers 
\cite{Page76,BelMar99,BelCac2010,WinYamA,WinYamB,BSa,BSb,BSW,SufFacCos83,FKSYinCPAM2000a,FKSYinCPAM2000b,FKSYinCMP2002,FKSYinATMP2003} 
in particular deal with Dirac's equation on some member of the Kerr--Newman family of spacetimes.
 However, to the best of our knowledge, nobody has yet investigated Dirac's equation on the entire
maximal analytically extended, topologically nontrivial Kerr--Newman spacetime.
 Such an investigation faces many conceptual and technical obstacles, but it becomes feasible 
in a zero-gravity limit which preserves the nontrivial topology of the Kerr--Newman spacetime 
and its associated electromagnetic structures. 
 In this limit one can rigorously study these general-relativistic effects on the Dirac Hamiltonian, 
separated from --- and not obscured by --- those caused by general-relativistic gravity.
 The results of such a zero-gravity investigation \cite{TZzGKN,KTZzGKNDa} are 
reported here.

 Readers whose expertise includes hyperbolic partial differential equations on nontrivial background 
spacetimes, and who right away want to find out about the results that we have obtained, may now want 
to jump to the technical section 3.
 Readers with expertise elsewhere in mathematical physics may find the few introductory lines 
written above hardly motivating enough to read on, however.
 Fortunately, a study of Dirac's equation on a zero-gravity Kerr--Newman spacetime 
can be motivated in at least two different other ways, one of which we are going to elaborate on 
in the next section. 
 There we discuss the \emph{perplexing problem} of the general-relativistic Hydrogen spectrum,
which ought to be interesting to most mathematical quantum physicists.\footnote{We are not
   suggesting that experimental physicists should worry about this academic problem.
   For the empirically relevant problem to estimate the influence of, say, Earth's gravitational field on
   the spectrum of Hydrogen in the lab, see Papapetrou \cite{Pap56}.}
 Yet another way to motivate our study --- which is even more intriguing, but was not yet ready for
public announcement at the time of the Regensburg conference and is, therefore, only briefly
mentioned here (in the last section) --- has meanwhile been made public in our paper \cite{KTZzGKNDb}.

 Our results are stated informally in the context of the general-relativistic Hydrogen problem at the 
end of the next section, while the precise statements and some technical details are given in section 3.
 In section 4 we list open questions left unanswered by our study, and we indicate 
the key idea of \cite{KTZzGKNDb}.
\vspace{-7pt}

\section{On the General-Relativistic Hydrogen Spectrum}

 Hydrogen  has played a crucial role in the development of the quantum theory of atomic spectra, and presumably 
this simplest of the chemical atoms will continue to play an important role in the ongoing efforts to find a more
satisfactory theory; for instance, one that does not rely on artificial UV cutoffs, etc.
 Yet we do not have to venture into the realm of quantum field theory or quantum gravity to encounter perplexing 
issues that await clarification.
 We simply ask for a general-relativistic counterpart of the special-relativistic spectrum of the quantum mechanical 
Dirac Hamiltonian for an electron (modeled as a  point charge) in the electromagnetic field of a proton (modeled 
either as a point --- or spherical  --- charge, or as a combination of electric charge plus current
distribution to account also for the proton's magnetic dipole field).
 Since {gravity is very weak} one would expect the general-relativistic Dirac point spectrum 
to differ from Sommerfeld's fine structure formula only by the tiniest amounts,
and in particular to be computable perturbatively using Newton's constant of universal gravitation, $G$, 
as expansion parameter. 
 	But if that is indeed what one expects, then one will be in for a surprise!

\subsection{The Coulomb Approximation}

 In this subsection only the electric field of the proton is considered. 
 To have a reference point, we begin by recalling the spectral results of the
familiar textbook problem, which is Dirac's equation for a point electron in flat Minkowski spacetime equipped
with a proton, modeled as a point charge having straight worldline; in the rest frame
of this model proton the electron experiences only the electrostatic Coulomb field of 
the point charge.
 Subsequently we turn to the general-relativistic version of this problem.

\subsubsection{\hspace{-2pt}The Special-Relativistic Spectrum}

 The pertinent Dirac Hamiltonian with domain\footnote{We follow the notation of
 Lieb and Loss \cite{LiebLossBOOK}; thus  $C^\infty_c(\Rset^3\backslash\{\boldsymbol{0}\})$ denotes functions which
are compactly supported away from the origin in $\Rset^3$.}
$C^\infty_c(\Rset^3\backslash\{\boldsymbol{0}\})^4$
is essentially self-adjoint on $L^2(\Rset^3\backslash\{\boldsymbol{0}\})^4$
with spectrum $\sigma = \sigma_{\mbox{\tiny{ac}}} \cup \sigma_{\mbox{\tiny{pp}}}$, the
 absolutely continuous part of which is given by
\begin{equation}
\sigma_{\mbox{\tiny{ac}}} 
= (-\infty,-m]\cup[m,\infty),
\end{equation}
where $m$ is the empirical mass of the electron, while the discrete (here equal to the pure point) part is
given by Sommerfeld's famous fine structure formula
\begin{equation}\begin{aligned}
\sigma_{\mbox{\tiny{pp}}} 
&= \label{sigmaSOMMERFELD}
\Biggl\{ 
mc^2\Biggl({1+\frac{\alpha_{\mbox{\tiny{S}}}^2}{\bigl(n-\kappa+\sqrt{\kappa^2-\alpha_{\mbox{\tiny{S}}}^2}\,\bigr)^2}}\Biggr)^{-1/2}
  \Biggr\}_{\genfrac{}{}{0pt}{}{\!n=1,...,\infty}{\!\kappa=1,...,n}} \\
&= \left\{ mc^2 \left(1- \frac{ \alpha_{\mbox{\tiny{S}}}^2}{2n^2}\right)\right\}_{n\in\Nset} + mc^2 O(\alpha_{\mbox{\tiny{S}}}^4),
\end{aligned}
\end{equation}
where $\kappa=j+1/2$, with $j\in \{1/2,...,n-1/2\}$ being nowadays total angular momentum quantum 
number.\footnote{For a modern semi-classical approach that produces these quantum numbers, see \cite{KeppelerBOOK}.}
 In (\ref{sigmaSOMMERFELD}), the expansion of the discrete spectrum in powers of Sommerfeld's fine structure constant 
$\alpha_{\mbox{\tiny{S}}}^{} = \frac{e^2}{\hbar c} \approx \frac{1}{137.036}$ reminds us that, except for the constant 
shift\footnote{The additive constant $mc^2$ drops out in the calculation of Rydberg's empirical formula for the 
  frequencies of the emitted / absorbed radiation, which are proportional to the  {differences} of the discrete 
  energy eigenvalues.}
by the electron's rest energy $mc^2$, 
special relativity only makes tiny corrections  $mc^2\times O(\alpha_{\mbox{\tiny{S}}}^4)$ 
to the Born--Oppenheimer approximation $m_{\mbox{\tiny p}}\to\infty$ of Bohr's energy spectrum
\begin{equation}
\sigma_{\mbox{\tiny{pp}}}^{\mbox{\tiny{Bohr}}}
 = \left\{- \frac{\mu c^2 \alpha_{\mbox{\tiny{S}}}^2}{2n^2}\right\}_{n\in\Nset}
\end{equation}
$\mu = \frac{mm_{\mbox{\tiny p}}}{m+m_{\mbox{\tiny p}}}$ is the reduced mass of the Hydrogen atom, so
$\mu\to m$ as $m_{\mbox{\tiny p}}\to\infty$.

\subsubsection{The Dirac Electron in Reissner--Nordstr\"om Spacetime}

\hspace{-3.5pt}
 We next switch on $G$ and ask for the general-relativistic spectrum of a ``test'' electron 
in the Reissner--Nordstr\"om ``electromagnetic spacetime of a point proton.''
 This spacetime is a spherically symmetric, eventually (in an open neighborhood of spacelike infinity) static, charged
solution of the Einstein--Maxwell equations (see below),
having a metric $\bg$ with line element $ds_\bg^2 = g_{\mu\nu}dx^\mu dx^\nu$ given by
\begin{equation}\label{RNdsSQ}
ds_\bg^2 =  {f(r)}c^2dt^2  - {f(r)^{-1}} dr^2 + r^2 (d\theta^2 + \sin^2\theta d\varphi^2),
\end{equation}
\begin{equation}
f(r)\equiv \left(1-\frac{2Gm_{\mbox{\tiny p}}}{c^2 r}+ \frac{Ge^2}{c^4 r^2}\right) ;
\end{equation}
here, $(t,r,\theta,\varphi)$ are Schwarzschild-type coordinates which asymptotically near spacelike infinity become just
the spherical coordinates of Minkowski spacetime (obtained by setting $f(r)\equiv 1$ in the above metric). 
 For the empirical values of $m_{\mbox{\tiny p}}$ and $e$, one has $Gm_{\mbox{\tiny p}}^2/e^2 \ll 1$, so
this spacetime is then static everywhere and covered by a single chart of $(t,r,\theta,\varphi)$ coordinates, 
exhibiting a timelike \emph{Naked Singularity}\footnote{This well-known
  naked singularity is usually not considered to be a counterexample of the (weak) \emph{cosmic censorship hypothesis}, 
  based on the following reasoning: paraphrasing Freeman Dyson, general relativity is a classical physical theory 
which applies \emph{only to physics in the large} (e.g. astrophysical and cosmic scales), not to atomic physics; 
and so, since cosmic
  bodies of mass $\textsc{m}$ and charge $\textsc{q}$ \emph{must} have a ratio $G\textsc{m}^2/\textsc{q}^2 \gg 1$, 
the Reissner--Nordstr\"om spacetime of 
  such a body (assumed spherical), when collapsed, exhibits a black hole, not a naked singularity. 
    While we agree that cosmic bodies (in mechanical virial-equilibrium) must have a ratio $G\textsc{m}^2/\textsc{q}^2 \gg 1$, 
  we don't see why the successful applications of general relativity theory at astrophysical and cosmic scales would
  imply that general relativity cannot be successfully applied at atomic, or even sub-atomic scales, where typically
$G\textsc{m}^2/e^2 \ll 1$.}
at $r=0$. 

  In the naked singularity sector  $Gm_{\mbox{\tiny p}}^2 <e^2$ (recall that the empirical proton's (mass, charge) pair belongs in 
this sector) one is confronted with the perhaps unexpected result that the Dirac Hamiltonian is not essentially self-adjoint --- 
any general relativist who abhors naked singularities will presumably feel vindicated by this result.
 Yet, as shown in \cite{CohPow}, \cite{Bel98}, \cite{BMB00}, there exists a one-parameter family of self-adjoint extensions 
of the Dirac operator with domain $C^\infty_c(\Rset^3\backslash\{\boldsymbol{0}\})^4$ which commute with the angular momentum 
operator, and all of these have an absolutely continuous spectrum given by
\begin{equation}
\sigma_{\mbox{\tiny{ac}}} = (-\infty,-mc^2]\cup [mc^2,\infty);
\end{equation} 
furthermore, Cohen and Powers \cite{CohPow} show that any pure point spectrum can only be located inside the gap of the continuum.
 Unfortunately, Cohen and Powers merely state that their preliminary studies indicate the existence of eigenvalues, and
we are not aware of any work that has actually shown the existence of eigenvalues for any of the self-adjoint extensions
of the formal Dirac operator on the naked Reissner--Nordstr\"om spacetime.\footnote{Interestingly enough, though, 
Belgiorno--Martellini--Baldicchi \cite{BMB00} proved the existence of bound states 
of a Dirac point electron \emph{equipped with an anomalous magnetic moment}
in the Reissner--Nordstr\"om spacetime with naked singularity, provided the anomalous magnetic moment is large enough;
in that case, the Dirac Hamiltonian is essentially self-adjoint.}

\begin{rem} 
One may also want to replace $m_{\mbox{\tiny p}}$ by other positive values, in particular by
$m_{\mbox{\tiny D}}\approx 2m_{\mbox{\tiny p}}$ (to study the Deuterium spectrum) and by 
$m_{\mbox{\tiny T}}\approx 3m_{\mbox{\tiny p}}$ (to study the Tritium spectrum).
 These choices leave one in the naked-singularity sector of the Reissner--Nordstr\"om spacetimes.
\hfill ${\qed}$
\end{rem} 
\vfill

 While the general-relativistic Hydrogen, Deuterium, and Tritium problems formulated with the Reissner--Nordstr\"om spacetime
for an electrostatic point proton inevitably lead to the naked singularity sector, 
mathematical physicists have also studied a whole family of ``hydrogenic problems'' with other 
possible positive mass values $\textsc{m}$ in place of $m_{\mbox{\tiny p}}$. 
 When $G\textsc{m}^2 \geq e^2$, the analytical extension of the outer Reissner--Nordstr\"om spacetime will feature
the \emph{Event Horizon} of a \emph{Black Hole} behind which lurks the timelike singularity.
 Moreover, when  $G\textsc{m}^2 > e^2$ then there exists yet another, inner horizon between the 
timelike singularity and the event horizon, and the region between this and the event horizon is not static. 
 Furthermore, the maximal analytical extension features multiple copies of these spacetime patches, which to some
extent are causally separated by \emph{Cauchy Horizons}. 

 For the black hole sector $G\textsc{m}^2>e^2$  of parameter space Cohen and Powers \cite{CohPow}
showed that the Dirac Hamiltonian is essentially self-adjoint on the set of $C^\infty$ bispinor-valued 
functions which are compactly supported outside the event horizon, but its spectrum is the \emph{whole real line}, 
and the pure point spectrum is empty. 
 The problem was picked up again by Belgiorno \cite{Bel98} and by Finster et al. \cite{FSYinJMP2000}, 
who also proved the absence of bound states supported outside the outer event horizon 
in the Reissner--Nordstr\"om black hole spacetime. 
 The latter authors considered also bispinor wave functions which are supported on both sides of the event horizon; in
particular, they also showed that in the extreme case $G\textsc{m}^2=e^2$ any bound state must be supported entirely 
behind the event horizon.
 An interesting open question is whether in the subextreme black hole sector $G\textsc{m}^2> e^2$ 
any bound state of a self-adjoint Dirac operator must be supported entirely inside the event horizon.

 Now, according to the mainstream view of relativists, only the black hole sector of a spacetime family 
is physically relevant, and for physicists taking the (for a long time also mainstream) positivistic view 
only the part outside of the black hole's event horizon is of concern to physics --- 
\emph{this combination of viewpoints thus forces one to conclude that in such a physical 
Reissner--Nordstr\"om spacetime there are no bound states of a Dirac point electron without 
anomalous magnetic moment.}

 But positivism is just a form of philosophy, not universally shared by all physicists.
 And so, if with Werner Israel one believes --- as we do --- that general relativity makes statements about the
\emph{physics} inside the event horizon of a black hole, and explores mathematically what it says the physics is,
then \emph{bound states of a Dirac point electron without anomalous magnetic moment may conceivably
exist in a physical Reissner--Nordstr\"om black hole spacetime, namely supported inside the event horizon.}
 In the same vein we may as well ignore the 
censorship hypothesis for the naked-singularity spacetimes and worry about whether
\emph{bound states of a Dirac point electron without anomalous magnetic moment exist in a Reissner--Nordstr\"om spacetime 
with naked singularity.} 
 In either case their existence would yet have to be proved.

 But one thing seems clear: none of these putative point spectra can be obtained perturbatively
from Sommerfeld's fine structure spectrum by ``switching on $G$.''
 In particular, the black hole point spectrum would presumably vanish as $G\to 0$ because the black hole itself 
vanishes in this limit, and so bear no resemblance to Sommerfeld's fine structure spectrum as $G\to 0$.
 And whether any part of any of the hypothetical point spectra for the naked singularity sector will 
resemble Sommerfeld's fine structure spectrum as $G\to 0$ is anybody's best guess.
 Hopefully someone will work it out eventually!

\begin{rem}
 The putative failure of $G$-perturbative reasoning can be traced to the
non-integrable electromagnetic stresses and energy density which are the source terms for
the Ricci curvature tensor of the Reissner--Nordstr\"om spacetime.
 This is the old problem of infinite electrostatic self-energy of a point charge, which 
because of the equivalence of energy and mass becomes its infinite self-mass problem.
 In a special-relativistic setting it assigns an infinite inertia to a point charge, which hounds one when trying to 
formulate a dynamical theory of charged point particle motion beyond the test particle approximation, but in 
a general-relativistic setting the gravitational coupling leads, in addition, to very strong curvature singularities of 
the spacetime generated by the non-integrable self-energy(-etc.) densities.
 Interestingly enough, for the remarkably accurate computation of the special-relativistic quantum-mechanical point 
spectrum of hydrogen (in the Born--Oppenheimer approximation) only the electrostatic interaction energy of a point proton
and a point electron enters the Dirac equation through the usual ``minimal coupling,'' i.e. both self-energy terms are ignored. 
 These self-energy terms are also ignored in the electromagnetic minimal coupling term of Dirac's equation on a 
Reissner--Nordstr\"om spacetime.
 But the electric self-energy density of the point proton enters the general-relativistic Dirac equation of a point electron
in the Reissner--Nordstr\"om spacetime also through the covariant derivatives of the spacetime, for
this non-integrable density is a curvature source term in the Einstein--Maxwell equations.
 So one may contemplate purging it, too. 
 This leads to the vacuum Einstein equations, and so instead of the Reissner--Nordstr\"om spacetime 
one would obtain the Schwarzschild spacetime; see \cite{HeineckeHehl} for a recent pedagogical treatment.
 Yet to retain the electrostatic interaction between point electron and proton
in the Dirac Hamiltonian, consistent with this approximation one would next have to solve the
Maxwell equations with a point proton as source in a Schwarzschild background spacetime, with the point proton
located in its ``center'' --- if this makes any sense at all --- and then treat the Dirac point electron as a test
charge experiencing the uncharged background metric as well as the electric field of the point proton imposed on 
that background metric.
 We are not aware of any such study; furthermore, we are not sure whether a mathematically well-posed formulation of
the indicated classical electrical problem is feasible because the Schwarzschild black hole spacetime is not static 
inside its event horizon, and its singularity is spacelike, so that its ``center'' is a spacelike
line, not a point, raising the question where exactly to place the point proton!
 Yet it may be mathematically interesting to sort this out.
\hfill ${\qed}$
\end{rem}

 We close this subsection by emphasizing that the just contemplated removal of the self-energy density of the point proton 
from the spacetime equations is a rather contrived step and not easily justifiable --- if at all ---, in contrast to the 
readily vindicated omission of the infinite \emph{self-interaction} terms from the \emph{electromagnetic energy} 
in the Hamiltonian.
 It has the flavor of ``a last desperate attempt'' to cling to
the point proton approximation when setting up the general-relativistic Hydrogen problem for a Dirac ``test electron'' 
interacting with it.
 Since a physical proton is a compound particle with a finite size, 
the mathematical catastrophes associated with the point proton approximation may well be dismissed 
as the result of an oversimplification and declared not a cause for real concern. 
 Indeed, since it is known that finite-size proton models remove the special-relativistic catastrophe of the hydrogenic
problem at $\textsc{z} = 1/\alpha_{\mbox{\tiny{S}}}\approx 137.036$ \cite{GreinerETalBOOK}, it is not difficult to
convince oneself that a finite-size model of a spherical proton avoids the general-relativistic spacetime singularity 
of the Reissner--Nordstr\"om spacetime of a point proton. 
 Although this introduces the problem of having to make assumptions about the structure of the proton, the tiny 
size of the proton suggests that all possible spherical finite size models should yield the same leading order 
corrections (in terms of powers of $G$) to the special-relativistic spectrum. 
 For example, assuming a model that produces a spherical surface charge, and zero binding energy,
one would obtain a spacetime which coincides with the Reissner--Nordstr\"om spacetime for $r> r_{\mbox{\tiny p}}$, 
and which is flat for $r<r_{\mbox{\tiny p}}$. 
 Here, $r_{\mbox{\tiny p}}$ is the solution to 
 $m_{\mbox{\tiny p}}c^2 - {e^2}/{r}=0$, viz. 
 $r_{\mbox{\tiny p}} = {e^2}/{m_{\mbox{\tiny p}} c^2}$, 
which is $\approx 1836$ times smaller than the so-called ``classical electron radius'' 
 ${e^2}/{m c^2}$,
where $m$ is the electron's empirical mass.
 The spacetime is not smooth at $r_{\mbox{\tiny p}}$ but its singularity corresponds to just a jump in its
Ricci curvatures. 
 So a $G$-perturbative calculation of the Dirac spectrum should be feasible.

\subsection{The Hyperfine Structure}

 So far we have assumed that the proton has only an electric charge.
 However, the physical proton appears to also possess a magnetic dipole moment. 
 The interaction of the electron spin-magnetic moment with this  magnetic moment of the proton accounts for
a hyperfine structure of the Hydrogen spectrum, as computed with quantum-mechanical perturbation theory.
 Unfortunately, assuming a point proton carrying an electric charge and magnetic dipole is QM-non-perturbatively catastrophic
even in a non-relativistic setting.
 A QM-non-perturbative calculation requires a model of a finite-size proton. 
 Pekeris \cite{Pek87} proposed that as a substitute for such a finite-size model of the proton with charge and currents
one may want to take the well-known electromagnetic Kerr--Newman spacetime family with its ring singularity and electromagnetic 
fields which, near spacelike infinity, approach an electric monopole and a magnetic dipole structure.
 Of course, this proposal should not be taken too seriously, in the sense that the inner structure of the proton is hardly 
 reproduced correctly by the Kerr--Newman solution.
 Yet it is certainly interesting
to investigate Dirac's equation for a point electron in the Kerr--Newman spacetime with its parameters matched to those of 
the proton.

\subsubsection{The Dirac Electron in the Kerr--Newman Spacetime with $G>0$}

 In the spirit of the previous subsection we now inquire into the general-relativistic spectrum 
of a ``test'' electron in the electromagnetic Kerr--Newman spacetime \cite{NewmanETal}, pretending that its
electromagnetic fields represent those of an extended proton with charge and magnetic moment.
 We do not display its metric (the line element of which is much more complicated than \eqref{RNdsSQ}; 
a special case will be exhibited further below, though) but only mention that it has three parameters --- 
charge $\textsc{q}$ (here chosen to equal $e$), ADM mass $\textsc{m}$ (here chosen to equal $m_{\mbox{\tiny p}}$), 
and ADM angular momentum per unit mass, $a$, here to be chosen such that $ea$ equals the magnetic moment of the proton. 
 This puts us into the naked singularity sector, but as before, we give a mini-survey 
of both, naked singularity sector and black hole sector, cf. \cite{Car68}.

\begin{itemize}
\item The open black hole sector, $G\textsc{m}^2 > {\textsc{q}^2+a^2c^4/G}$,
was studied in \cite{BelMar99, FKSYinCPAM2000a, FKSYinCPAM2000b, WinYamB}, and
no bound states of the Dirac equation were found for its domain supported either outside the
event horizon or on both sides of it;\footnote{The addition of a positive cosmological constant \cite{BelCac2010} 
	has not lead to bound states either.}
the latter situation was studied only in \cite{FKSYinCPAM2000a, FKSYinCPAM2000b}, involving
also a matching across the Cauchy horizon lurking inside the event horizon.


\item Interestingly, in the extreme case  $G\textsc{m}^2={\textsc{q}^2+a^2c^4/G}$ (the boundary of the open 
Kerr--Newman black hole sector, which belongs to the black hole sector, too), bound states 
supported outside the event horizon exist for a sequence of special $m$ values \cite{WinYamB}.

\item In the naked singularity sector, $G\textsc{m}^2 < {\textsc{q}^2+a^2c^4/G}$, the whole spacetime manifold is causally vicious, and 
we are not aware of any study of the Dirac equation on it.
\end{itemize}

 Since in this subsection we inquire into whether a $G$-deforma\-tion of the Sommerfeld spectrum with 
hyperfine corrections can be computed by studying Dirac's equation on the Kerr--Newman spacetime,
the outcome is somewhat disappointing: the proton parameters $\textsc{m} = m_{\mbox{\tiny p}}$ and $\textsc{q}=e$ 
belong to the naked singularity sector of the Kerr--Newman family, 
and nothing seems to be known about the Dirac equation on it.

 On the other hand, since the proton mass does not enter the Sommerfeld fine structure formula, one may still ask 
about the $G$-dependence of the discrete Dirac spectra in the extreme Kerr--Newman black hole spacetime and whether they
resemble the Sommerfeld spectrum with hyperfine corrections as $G\to 0$; unfortunately, so far not much is known about these 
point spectra, either, but  someone should work out the answer eventually!
 
\subsubsection{The Dirac Electron in the zero-$G$ Kerr--Newman Spacetime}

 The inquiry started in the previous subsection suggests a closely related spectral question about the Dirac operator and the
Kerr--Newman spacetime, one which avoids all the causal pathologies associated with the latter. 
 Namely, since the canonical (in the sense of Geroch \cite{Geroch}) zero-$G$ limit of the maximal analytically extended 
Kerr--Newman spacetime (z$G$KN) does \emph{not} yield the Minkowski spacetime but a flat yet topologically nontrivial spacetime 
with a ring singularity \cite{TZzGKN}, it is an interesting question whether the Dirac spectrum for an electron in this spacetime 
bears any resemblance  to the Sommerfeld spectrum with hyperfine corrections.

\begin{figure}[ht]
\begin{center}
\includegraphics[scale=0.45]{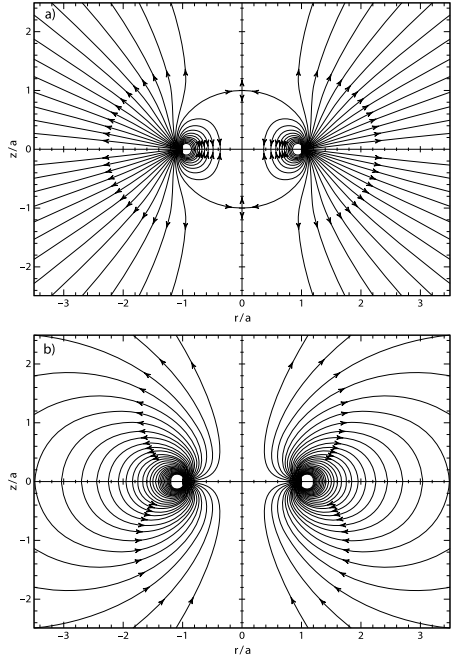}
\caption{(From \cite{GairPRIZE} and \cite{LB})
Electric (top) and magnetic (bottom) lines of force in a Euclidean plane containing the 
($z$-)axis of symmetry of a constant-$t$ section of a \emph{single sheet} of the z$G$KN spacetime; 
note that in these plots, $r$ denotes a Cartesian coordinate $\perp z$, and not
the Boyer--Lindquist coordinate!
  The orientation of the lines (indicated by the arrows) reverses across the straight line segment
  between the two singular points (which are located at $(r/a,z/a)=(-1,0)$ and $(r/a,z/a)=(1,0)$); associated 
with this reversal is a jump discontinuity in the magnitudes of the respective field strengths when 
one crosses that line segment.\label{fig:KNfieldsEandB}}
\end{center}
\end{figure}

 The apparently first investigation in this direction is by Pekeris \cite{Pek87}.
 However, following Israel \cite{Israel}, he works with the zero-$G$ limit of a \emph{single sheet} of the Kerr--Newman
electromagnetic spacetime, which is a Minkowski spacetime decorated with truncated multi-valued harmonic fields.
 Fig.~\ref{fig:KNfieldsEandB}, produced by J. Gair and published in \cite{GairPRIZE}, and in
\cite{LB}  by D. Lynden-Bell, and which is reproduced here with permission from both D. Lynden-Bell and J. Gair, shows a drawing of electric 
(top) and magnetic (bottom) lines of force in a planar section, containing the axis of symmetry, of a spacelike snapshot.
 The ring singularity pierces the drawing at the two singular points where all the field lines seem to emerge from,
respectively end at. 

 If one chooses to interpret the zero-$G$ limit of the spacetime in this single-sheeted way, then one is forced to 
interpret the inevitable jump discontinuities in the electromagnetic fields as being caused by ultra-singular 
two-dimensional sources. 
 A geometrically distinguished choice of such a source is the ultra-singular disc source
spanned by the ring singularity, studied by  \cite{Israel}, \cite{Pek87}, \cite{GairPRIZE}, \cite{LB}, and \cite{Kaiser}. 
  These disc-type charge and current densities are not integrable, but are magically compensated in parts by oppositely infinite 
charges and currents on the ring, in such a manner that the finite charge of the Kerr--Newman fields is produced.

 By contrast, from the perspective of a two-sheeted interpretation of the z$G$KN spacetime and its 
electromagnetic fields the jump discontinuities across the line spanned by the two singular points 
seen in Fig.~\ref{fig:KNfieldsEandB} are artifacts of the \emph{single-sheeted drawing} of the 
multi-valued harmonic functions with branch cut placed arbitrarily at the disc spanned by the singular ring.
 Namely, the sources of the fields living on the double-sheeted maximal analytically extended zero-$G$
Kerr--Newman spacetime are finite sesqui-poles concentrated in the singular ring, see \cite{TZzGKN}.
 Thus the Dirac equation on this maximal analytically extended zero-$G$ Kerr--Newman spacetime can be studied in
an orderly manner. 
 We have begun such a investigation \cite{KTZzGKNDa,KTZzGKNDb} of Dirac's equation on the maximal analytically extended 
zero-$G$ Kerr--Newman spacetime, and in the following we report on it.

 In the zero-$G$ limit of the maximal analytically extended Kerr--Newman spacetime with metric expressed in Boyer--Lindquist 
coordinates $(t,r,\theta,\varphi)$, one obtains a flat double-sheeted spacetime $\cM$ with Zipoy topology \cite{Zipoy}, having a 
metric $\bg$ with line element $ds_\bg^2 = g_{\mu\nu}dx^\mu dx^\nu$ given by
\begin{equation}\label{zGKNg}
ds_\bg^2 = c^2dt^2 -  \frac{r^2+a^2\cos^2\theta}{r^2+a^2}(dr^2 + (r^2+a^2)d\theta^2) - (r^2+a^2)\sin^2\theta d\varphi^2;
\end{equation}
here, $-\infty<t<\infty$, $-\infty<r<\infty$, $0\leq \theta\leq \pi $, $0\leq \varphi < 2\pi$.
 By $\mathcal{N}$ 
we denote any of the spacelike $t=const.$ slices of the z$G$KN spacetime $\cM$;
note that $\mathcal{N}$ is independent of $t$.
 The z$G$KN electromagnetic field is an exact two-form, $\bF = d\bA$, with  
\begin{equation}\label{zGKNA}
\bA = -\frac{r}{r^2 + a^2 \cos^2\theta}(\textsc{q}cdt - \textsc{q}a \sin^2\theta d\varphi).
\end{equation} 
 We have studied Dirac's Hamiltonian for an electron in the above electromagnetic spacetime.
 An informal summary of our main results follows:

\begin{itemize}
\item
 The Dirac operator with domain $C^\infty_c(\mathcal{N},\Cset)^4$ is {essentially self-adjoint}.
\item
 Its unique self-adjoint extension has a {symmetric} spectrum about zero.
\item
 Its continuous spectrum is given by $\sigma_{ac} = (-\infty,-mc^2]\cup[mc^2,\infty)$.
\item
 Its discrete spectrum is {non-empty} if $2|a|{mc}/{\hbar}  < 1$ and also
$|e\textsc{q}|/\hbar c < \sqrt{(2|a|mc/\hbar) (1-2|a|{mc}/{\hbar})}$,
 and then it is located inside $(-mc^2,mc^2)$.
\item
 If the limit $a\to 0$ of the discrete spectrum converges to the spectrum of the Dirac Hamiltonian on the 
$a\to0$ electromagnetic spacetime, it is a union of a positive and a negative Sommerfeld fine structure spectrum.
\end{itemize}

\begin{rem}
 Our sufficient conditions for the existence of the discrete spectrum presumably are not necessary conditions.
 In any event, for our inquiry into general-relativistic effects on the hydrogen spectrum we use
the empirical magnetic moment  of the proton and find $2a\approx 10^{-3}\hbar/mc$ (in rough agreement with the
empirical size of the proton; cf. \cite{Pek87}), so the first sufficient smallness condition is fulfilled 
(note that it demands that the ring diameter $2|a|$ is smaller than the electron's Compton wavelength $\hbar/mc$).
 And for $\textsc{q}=e$ the l.h.s. of the second sufficient smallness condition becomes 
$\alpha_{\mbox{\tiny{S}}}^{} \approx 1/137.036$, so the second condition is fulfilled then, too.
\hfill ${\qed}$
\end{rem}

\section{\hspace{-7pt} A zero-$G$ Kerr--Newman Born-Oppenheimer Hydrogen Atom}

 In this section we present the essentials of our study.
 We begin by describing the zero-$G$ Kerr--Newman spacetime and its electromagnetic
field in more detail.
  Then we formulate Dirac's equation on it in its ``standard'' format, using Cartan's frame method,
which allows us to define the Dirac Hamiltonian on a static, spacelike slice of the z$G$KN spacetime.
 Finally, we state precisely our results, together with a few remarks regarding our proofs.

\subsection{The zero-Gravity Spacetimes and their Electromagnetic Fields}

\subsubsection{The Einstein--Maxwell Equations}
 
 An electromagnetic spacetime is a triple $(\cM,\bg,\bF)$, where $(\cM,\bg)$ is a four-dimensional Lorentz manifold with
metric $\bg$, and $\bF=d\bA$ is the Faraday tensor of the electromagnetic field on $\cM$.
 The Einstein--Maxwell equations for an electromagnetic spacetime are a system of PDE given by Einstein's field equations
\begin{equation}
R_{\mu\nu}[\bg] - \frac{1}{2} R g_{\mu\nu} =\frac{8\pi G}{c^4} T_{\mu\nu}[\bF,\bg],
\end{equation}
with $\mu\in\{0,1,2,3\}$ and $\nu\in\{0,1,2,3\}$, and where
\begin{equation}
T_{\mu\nu}[\bF,\bg] 
= \frac{1}{4\pi}\left( F_\mu^\lambda F_{\nu\lambda} - g_{\mu\nu} F_{\alpha\beta}F^{\alpha\beta} \right)
\end{equation} 
is the energy(-density)-momentum(-density)-stress tensor of the electromagnetic field $\bF$.
 The Bianchi identities $\nabla^\mu (R_{\mu\nu}[\bg] - \frac{1}{2} R g_{\mu\nu}) = 0$ imply the conservation laws 
\begin{equation}
\nabla^\mu T_{\mu\nu} = 0,
\end{equation}
which in turn imply that the Maxwell tensor $\bM=\star\bF$ satisfies  $d \bM =0$; here, $\star$ is the Hodge dual operator.
 Recall that $d\bF =0$ because $dd\bA=0$.

 Incidentally, the Einstein--Maxwell equations simplify somewhat due to the vanishing trace $T^\mu_\mu(\bF,\bg) =0$, 
which implies $R=0$.

\subsubsection{The zero-$G$ Kerr--Newman Spacetime and its Electromagnetic Field}

 The Kerr--Newman spacetime with its electromagnetic field is an axisymmetric, asymptotically flat and stationary, 
three-parameter solution of the above Einstein--Maxwell equations; see \cite{NewmanETal,Car68}.
 In the limit $G\to 0$ their spacetime metric becomes \eqref{zGKNg},
solving Einstein's vacuum equations $R_{\mu\nu} = 0$ --- usually obtained by 
setting $T_{\mu\nu} \equiv 0$ --- while their electromagnetic field $\bF=d\bA$, with $\bA$ given by \eqref{zGKNA}
solves the zero-$G$ Maxwell equations 
(which in our compact notation look unchanged) on the limiting zero-$G$ spacetime.
	
\begin{figure}[ht]
\begin{center}
\hspace{-.5cm}
\includegraphics[width=9cm,height=7cm]{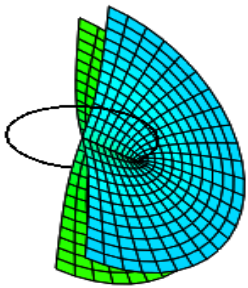}
\end{center}
\vspace{-1.5cm}
\caption{An illustration of the Zipoy topology.\label{fig:ZipTop}}
\end{figure}

 The z$G$KN spacetime is readily illustrated as follows. 
 Since it is static, it suffices to discuss a constant-$t$ snapshot, $\cN$, whose metric is given by the 
space part of \eqref{zGKNg}, with $(r,\theta,\varphi)$ oblate spheroidal coordinates. 
 Since $\cN$ is axisymmetric, it furthermore suffices to discuss a constant-$\varphi$ section of $\cN$.
 Shown in Fig.~\ref{fig:ZipTop}
  are the ring singularity and the part $\{r\in(-1,1),\theta\in(0,\pi)\}$ of such a constant-azimuth section of $\cN$
(slightly curved to separate the sheets for the purpose of visualization); the coordinate grid on the sheets shows the 
constant-$\theta$ lines (hyperbolas) and constant-$r$ lines (ellipses).
 Solutions to Einstein's equations having this two-sheeted topology were first discovered by Zipoy \cite{Zipoy}, 
for which reason we speak of Zipoy topology.
\vspace{-.4cm}$\phantom{nix}$

 To illustrate the z$G$KN electromagnetic field, we consider $\bE+i\bB = \mathbf{i}_{\partial_t^{}}(\bF+i\star\bF)$,
where $\bE$ and $\bB$ are the electric and magnetic fields, obtaining
\begin{equation}\label{AppellEB}
\bE+i\bB 
= - d \frac{\textsc{q}}{r - i a \cos\theta}.
\end{equation}
They were discovered in this form by Appell \cite{Appell} who realized that these are multi-valued harmonic fields on
Euclidean space. 
 The insight that multi-valued harmonic fields become single-valued on so-called 
{\em branched Riemann spaces} is due to Sommerfeld \cite{Som97}, whose pioneering work was generalized and completed
by Evans \cite{Evans} and his students.
 In particular, the fields $\bE$ and $\bB$ given in \eqref{AppellEB} are single-valued harmonic fields on $\cN$.
 Moreover, due to the axisymmetry, the lines of force of $\bE$ and $\bB$ are planar curves in doubled half-planes 
with Zipoy topology which contain the axis of symmetry.
 
\begin{figure}[ht]
\begin{center}
\includegraphics[scale=0.4]{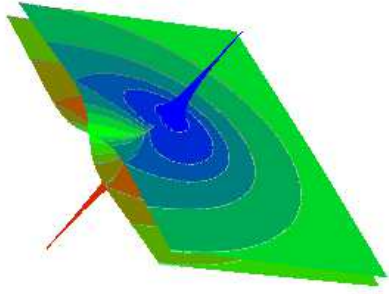}
\caption{Graph of the electric potential modulated onto a constant-$(t,\varphi)$ section of the z$G$KN spacetime.
The two sheets are separated and bend for purposes of visualization.\label{fig:zGKNpotE}}
\end{center}
\end{figure}

 As an illustration of the single-valuedness, and smoothness (except for their divergence at the ring
singularity) of the electromagnetic fields on the maximal analytically extended z$G$KN spacetime, in Fig.~\ref{fig:zGKNpotE}
we show the graph of the electric potential modulated onto the constant-$(t,\varphi)$ section of spacetime depicted in 
Fig.~\ref{fig:ZipTop}.
 The electric potential is positive on the upper and negative on the lower sheet, diverging at the ring singularity
(omitted in this picture, yet discernible due to the spikes in the potential), and smoothly ``criss-crossing'' at the 
disc spanned by the singular ring (a line in this constant-$(t,\varphi)$ section).
 The electric lines of force (not shown in Fig.~\ref{fig:zGKNpotE}) are orthogonal to the equipotentials.

\subsection{Relativistic Quantum Mechanics in z$G$KN Spacetime}

\subsubsection{The General-Relativistic Dirac Equation}

The Dirac equation for a test electron of charge $-e$ and mass $m$ in an electromagnetic Lorentz manifold 
is a first-order system of PDE for a bispinor field  $\Psi: (\cM,\bg,\bF) \to \Cset^4$ given by
\begin{equation}\label{firstDIRACeq}
{\tilde\gamma}^\mu (-i \hbar \nabla_\mu  + \textstyle{\frac{1}{c}} eA_\mu) \Psi + mc \Psi = 0;
\end{equation}
here, the $A_\mu$ are defined by $\bA = A_\mu dx^\mu$, and
the $\tilde{\gamma}^\mu$ are $4\times{4}$ Dirac matrices satisfying the fundamental identity of a
Clifford algebra,
\begin{equation}\label{Clifford}
\tilde{\gamma}^\mu 
\tilde{\gamma}^\nu + \tilde{\gamma}^\nu \tilde{\gamma}^\mu =  2 g^{\mu\nu}{\boldsymbol{1}}_{4\times4},
\end{equation}
where $(g^{\mu\nu})$ is the inverse matrix to $(g_{\mu\nu})$.
 Moreover, $\nabla_\mu$ denotes the covariant derivative for the metric $\bg$.

\subsubsection{Dirac's Equation in Standard Form} 

 While concise, the above version of Dirac's equation is not necessarily convenient for analytical studies.
 It is more convenient to use Cartan's frame formulation \cite{BriCoh}, which expresses the $\tilde\gamma$ matrices in terms
of the standard $\gamma$ matrices of Minkowski spacetime, thus
\begin{equation}\label{DIRACmatricesCOMEin}
\tilde{\gamma}^\mu \nabla_\mu    = \gamma^\mu\be_\mu  +\frac{1}{4} \Omega_{\mu\nu\lambda} \gamma^\lambda\gamma^\mu\gamma^\nu;
\end{equation}
here we introduced Dirac's gamma matrices, satisfying
\begin{equation}\label{DIRACmatricesCLIFFORD}
\gamma^\nu\gamma^\mu+\gamma^\mu\gamma^\nu = 2{{\eta}}^{\mu\nu}{\boldsymbol{1}}_{4\times4} ,
\end{equation}
where
\begin{equation}
({\boldsymbol{\eta}}) = \diag(1,-1,-1,-1)
\end{equation}
is the matrix of the Minkowski metric in rectangular coordinates,
and we also introduced the Ricci rotation coefficients $\Omega^\mu_{\nu\lambda}$, defined by
\begin{equation}\label{OMdef}
d \be^\mu = \Omega^\mu_{\nu\lambda} \be^\lambda \wedge \be^\nu
\end{equation}
plus an anti-symmetry condition in the lower two parameters; moreover, we introduced the orthonormal frame field
\begin{equation}\label{emuDEF}
(\be_\mu)^\nu (\be_\lambda)^\kappa g_{\nu\kappa} = {{\eta}}_{\mu\lambda} .
\end{equation}
 Finally, we obtain Dirac's equation in standard form (temporarily setting $\hbar=1=c$),
\begin{equation}\label{DIRACeqCARTANformat}
\gamma^\mu \left(\be_\mu + \Gamma_\mu + i e  \tilde{A}_\mu\right)\Psi  + im\Psi = 0,\qquad 
\Gamma_\mu := \frac{1}{4} \Omega_{\nu\lambda\mu} \gamma^\nu\gamma^\lambda;
\end{equation}
here 
\begin{equation}
\tilde{A}_\mu  := (\be_\mu)^\nu A_\nu.
\end{equation}

\subsubsection{The Standard Form of Dirac's Equation on z$G\,$KN} 

 We begin by introducing a Cartan (co-)frame $(\boldsymbol{\omega}^\mu)_{\mu=0}^3$ for the cotangent
bundle \cite{CarMcL82}:
\begin{equation}
\boldsymbol{\omega}^0\! := \frac{\varpi}{|\rho|} (dt - a \sin^2\theta\, d\varphi),
\boldsymbol{\omega}^1\! := |\rho|d\theta,
\boldsymbol{\omega}^2\! := \frac{\sin\theta}{|\rho|} (-a dt + \varpi^2 d\varphi),
\boldsymbol{\omega}^3\! := \frac{|\rho|}{\varpi}dr
\end{equation}
with the abbreviations
\begin{equation}
\varpi := \sqrt{r^2 + a^2}, \quad \rho:= r + i a \cos\theta.
\end{equation}
Let us denote the oblate spheroidal coordinates $(t,r,\theta,\varphi)$ collectively by $(y^\nu)$.
  Let $g_{\mu\nu}$ denote the coefficients of the z$G$KN spacetime metric in oblate spheroidal coordinates,
i.e. $g_{\mu\nu} = \bg\Big(\frac{\partial}{\partial y^\mu},\frac{\partial}{\partial y^\nu}\Big)$.
     One easily checks that written in the $(\boldsymbol{\omega}^{\mu})$ frame, the spacetime line element is
\begin{equation}
ds_{\bg}^2 = g_{\mu\nu}dy^\mu dy^\nu = {{\eta}}_{\alpha\beta} \boldsymbol{\omega}^{\alpha}\boldsymbol{\omega}^{\beta}.
\end{equation}
 With respect to this frame the electromagnetic Sommerfeld potential $\bA = \tilde{A}_\mu \boldsymbol{\omega}^\mu$, with
\begin{equation}\label{KNelmagAtilde}
\tilde{A}_0 =  -\textsc{q}\frac{r}{|\rho|\varpi} ,\quad
\tilde{A}_1 = 0, \quad
\tilde{A}_2 = 0,\quad
\tilde{A}_3 = 0.
\end{equation}

  The frame of vector fields $({\be}_{\mu})$ is the {\em dual} frame to $(\boldsymbol{\omega}^{\mu})$,
yielding an orthonormal basis for the tangent space at each point in the manifold:
\begin{equation}
\be_0 = \frac{\varpi}{|\rho|} \partial^{}_t + \frac{a}{\varpi|\rho|} \partial^{}_\varphi,\;
\be_1 = \frac{1}{|\rho|}\partial^{}_\theta,\;
\be_2 = \frac{a\sin\theta}{|\rho|} \partial^{}_t + \frac{1}{|\rho|\sin\theta} \partial^{}_\varphi,\;
\be_3 = \frac{\varpi}{|\rho|} \partial^{}_r\;.
\end{equation}

  Next, the anti-symmetric matrix $\big(\Omega_{\mu\nu}\big) = \big({{\eta}}_{\mu\lambda}\Omega^\lambda_\nu\big)$ is computed to be
\begin{equation}
(\Omega_{\mu\nu}) = \left(\begin{array}{cccc}
0 & -C\boldsymbol{\omega}^0 - D\boldsymbol{\omega}^2 & D\boldsymbol{\omega}^1 - B\boldsymbol{\omega}^3 & -A\boldsymbol{\omega}^0- B \boldsymbol{\omega}^2 \\
& 0 & D\boldsymbol{\omega}^0 + F\boldsymbol{\omega}^2 &-E \boldsymbol{\omega}^1 - C\boldsymbol{\omega}^3  \\
&\textrm{(anti-sym)}& 0 & -B \boldsymbol{\omega}^0 - E\boldsymbol{\omega}^2 \\
& & & 0
\end{array}\right),
\end{equation}
with
\begin{equation}
A := \frac{a^2 r \sin^2\theta}{\varpi |\rho|^3},\,
B := \frac{a r \sin\theta}{|\rho|^3},\,
C := \frac{a^2 \sin\theta\cos\theta}{|\rho|^3},\, 
\end{equation}
\begin{equation}
D := \frac{a\cos\theta\varpi}{|\rho|^3},\,
E := \frac{r\varpi}{|\rho|^3},\,
F := \frac{\varpi^2\cos\theta}{|\rho|^3\sin\theta}.
\end{equation}

 With respect to this frame on a z$G$K spacetime the covariant derivative part of
the Dirac operator  can be expressed with the help of the operator
\begin{equation}
\fO := \tilde{\gamma}^\mu\nabla_\mu = \left(\begin{array}{cc} 0 & \fl'+\fm'\\ \fl+\fm &0 \end{array}\right),
\end{equation}
where
\begin{equation}
\fl :=
\frac{1}{|\rho|} \left(\begin{array}{cc} D_+ & L_- \\ L_+ & D_-\end{array}\right)
\end{equation}
and
\begin{equation}
\fl' :=
\frac{1}{|\rho|} \left(\begin{array}{cc} D_- & -L_- \\ -L_+ & D_+\end{array}\right),
\end{equation}
with
\begin{equation}\label{eq:DpmLpm}
D_\pm := \pm \varpi \partial^{}_r + 
\varpi \partial^{}_t + \frac{a}{\varpi} \partial^{}_\varphi,
\quad
L_\pm :=\partial^{}_\theta  \pm i \left(a \sin\theta\,\partial^{}_t+\csc\theta \partial^{}_\varphi\right),
\end{equation}
while (with ${}^*$ denoting complex conjugation)
\begin{equation}
\begin{aligned}
\fm &:=  \frac12\bigl[ (-2C+F+iB)\sigma^{}_1+(-A+2E+iD)\sigma^{}_3\bigr] \\
&\ = \frac{1}{2|\rho|} \left(\begin{array}{cc} \frac{r}{\varpi}+ \frac{\varpi}{{\rho^*}} &\cot\theta +  \frac{ia\sin\theta}{{\rho^*\
}}\\
\cot\theta + \frac{ia\sin\theta}{{\rho^*}} &  -\frac{r}{\varpi} - \frac{\varpi}{{\rho^*}}\end{array}\right)
\end{aligned}
\end{equation}
and
\begin{equation}
\fm': = \frac12\bigl[ (2C-F+iB)\sigma^{}_1+(A-2E+iD)\sigma^{}_3\bigr] = -\fm^\dagger,
\end{equation}
where ${}^\dagger$ denotes Hermitian adjoint, and where the $\sigma^{}_k$ are Pauli matrices, viz.
\begin{equation}
 \sigma^{}_1 = \left(\begin{array}{cc} 0 & 1\\ 1 & 0\end{array}\right),\quad
 \sigma^{}_2 = \left(\begin{array}{cc} 0 & -i\\ i &\ 0\end{array}\right),\quad
 \sigma^{}_3 = \left(\begin{array}{cc} 1 & \ 0\\ 0 & -1\end{array}\right).
\end{equation}

  The principal part of $|\rho|\fO$ has an additive separation property,
\begin{equation}\label{eq:DDprincipal}
\begin{aligned}
|\rho|\left(\begin{array}{cc} 0 & \fl'\\ \fl & 0\end{array}\right)
=
\left[
 \gamma^3 \varpi \partial^{}_r + \gamma^0\left(\varpi \partial^{}_t + \frac{a}{\varpi} \partial^{}_\varphi\right)\right]
 + \\
\hspace{7pt} \Bigl[ \gamma^1 \partial^{}_\theta + \gamma^2(a\sin\theta \partial^{}_t + \csc\theta\,\partial^{}_\varphi)\Bigr],
\end{aligned}
\end{equation}
where the coefficients of the two square-bracketed operators are functions of only $r$, respectively only $\theta$,
and the lower order term in $\fO$ can be transformed away, achieving exact separation for $|\rho|\fO$.
 Namely, setting
\begin{equation}
\chi(r,\theta) := \frac12 \log( \varpi {\rho^*}\sin\theta),
\end{equation}
it is easy to see that
\begin{equation}
\fm = \fl\chi,\qquad \fm' = \fl'{\chi^*}.
\end{equation}
 Define now the diagonal matrix
\begin{equation}\label{def:D}
\fD := \diag( e^{-\chi},e^{-\chi}, e^{-{\chi^*}}, e^{-{\chi^*}})
\end{equation}
and a new bispinor $\hat{\Psi}$ related to the original $\Psi$ by
\begin{equation}
\Psi = \fD \hat{\Psi},
\end{equation}
and denoting the upper and lower components of a bispinor $\Psi$ by $\psi_1$ and $\psi_2$ respectively, it now follows that
\begin{equation}
(\fl + \fm)\psi_1 =
(\fl + \fm)(e^{-\chi}\hat{\psi}_1) =
 e^{-\chi} \left[ \fl - \fl\chi + \fm\right]\hat{\psi}_1 =
 e^{-\chi} \fl\hat{\psi}_1,
\end{equation}
and similarly
\begin{equation}
(\fl'+ \fm')\psi_2 = e^{-{\chi^*}} \fl'\hat{\psi}_2.
\end{equation}
 Setting 
\begin{equation}
\fR := \diag(\rho,\rho,{\rho^*},{\rho^*})
\end{equation}
and noting that $|\rho|\fD^{-\dagger}\fD = \fR$ while $\fD^{-\dagger}\gamma^\mu\fD = \gamma^\mu$ (where $\fD^{-\dagger}$ 
is shorthand for $(\fD^{-1})^\dagger$),
we insert $\Psi = \fD \hat{\Psi}$ in \eqref{DIRACeqCARTANformat} and left-multiply the equation by the diagonal matrix
$\fD' := |\rho|\fD^{-\dagger}$, and conclude that $\hat{\Psi}$ solves the transformed Dirac equation (with $\hbar=1=c$)
\begin{equation}\label{eq:newDir}
\left(|\rho|\gamma^\mu (\be_\mu + i e\tilde{A}_\mu) + im\fR\right) \hat{\Psi} = 0.
\end{equation}

\subsubsection{The Dirac Hamiltonian on a contant-$t$ Snapshot of z$G$KN Spacetime}
 We now recast Dirac's equation \eqref{eq:newDir} for $\Psi$ in Schr\"odinger form, 
\begin{equation}\label{eq:DIRACeqHAMformat}
i \hbar \partial^{}_t \hat{\Psi} = \hat{H}\hat{\Psi},
\end{equation}
for which we have to compute the Dirac Hamiltonian from \eqref{eq:newDir}. 
  Let matrices $M^\mu$ be defined by
\begin{equation}
|\rho|\gamma^\mu\be_\mu = M^\mu\partial^{}_\mu.
\end{equation}
  In particular,
\begin{equation}
M^0 = \varpi\gamma^0 + a \sin\theta\, \gamma^2.
\end{equation}
 We may thus rewrite \eqref{eq:newDir} as
\begin{equation}
M^0 \partial^{}_t\hat{\Psi} = - \left( M^k\partial^{}_k + ie|\rho|\gamma^\mu\tilde{A}_\mu + im\fR\right)\hat{\Psi}.
\end{equation}
 Finally, restoring $\hbar$ and $c$, and defining
\begin{equation}\label{def:Hhat}
\hat{H} := (M^0)^{-1} \left( M^k (-i \hbar \partial^{}_k) + {\textstyle\frac1c} e|\rho|\gamma^\mu\tilde{A}_\mu + mc\fR\right),
\end{equation}
we arrive at \eqref{eq:DIRACeqHAMformat}.

\subsubsection{A Hilbert Space for $\hat{H}$}\label{sec:Hilbert}

In order to obtain the correct inner product for the space of bispinor fields defined on the z$G$KN spacetime, 
we left-multiply the original Dirac equation \eqref{firstDIRACeq} by the conjugate bispinor $\overline{\Psi}$, defined as
\begin{equation}
\overline{\Psi} := \Psi^\dag \gamma^0,
\end{equation}
integrate the result over a slab of spacetime, and obtain the action for this equation.
 We find
\begin{equation}
\cS[\Psi]=\int_I dt \int_{\Sigma_t} \Psi^\dag \gamma^0 \left[ \tilde{\gamma}^\mu \nabla_\mu \Psi + \dots \right] d\mu^{}_{\Sigma_t},
\end{equation}
where $I\subset\Rset$ is a finite interval, and
$d\mu^{}_{\Sigma_t}$ is the volume element of $\Sigma_t\equiv\cN$, any spacelike $t=$ constant slices of z$G$KN.
 Using oblate spheroidal coordinates, with $d\mu_\cN = |\rho|^2\sin\theta  d\theta d\varphi dr$, 
it follows that the natural inner product for bispinors on $\Sigma_t=\cN$ needs to be
\begin{equation}
\langle \Psi,\Phi\rangle = \int_{\cN} \Psi^\dag\gamma^0\tilde{\gamma}^0 \Phi d\mu^{}_{\cN}
= \int_0^{2\pi}\int_0^\pi \int_{-\infty}^\infty \Psi^\dag M \Phi |\rho|^2 \sin\theta d\theta d\varphi dr,
\end{equation}
with 
\begin{equation}
M := 
\gamma^0 \tilde{\gamma}^0 = \gamma^0 \be_\nu^0 \gamma^\nu = \frac{\varpi}{|\rho|} \alpha^0 + \frac{a\sin\theta}{|\rho|} \alpha^2.
\end{equation}
Here, $\alpha^2$ is the second one of the three Dirac alpha matrices in the Weyl (spinor) representation
\begin{equation}
\alpha^k = \gamma^0 \gamma^k = \left(\begin{array}{cc} \sigma^{}_k & 0 \\ 0 & -\sigma^{}_k\end{array}\right),\qquad k=1,2,3,
\end{equation}
and for notational convenience the $4\times4$ identity matrix has been denoted by
\begin{equation}
\alpha^0 =  \left(\begin{array}{cc} \boldsymbol{1}_{2\times2} & 0 \\ 0 & \boldsymbol{1}_{2\times2}\end{array}\right).
\end{equation}

Now, let $\Psi = \fD \hat{\Psi}$ and $\Phi = \fD \hat{\Phi}$, with $\fD$ as in \eqref{def:D}.
  Then we have
\begin{equation}
\langle \Psi,\Phi \rangle = \int_0^{2\pi}\int_0^\pi \int_{-\infty}^\infty \hat{\Psi}^\dag \hat{M} \hat{\Phi} d\theta d\varphi dr,
\end{equation}
where
\begin{equation}
\hat{M} := \alpha^0 + \frac{a\sin\theta}{\varpi} \alpha^2.
\end{equation}
The eigenvalues of $\hat{M}$ are $\lambda_\pm = 1 \pm \frac{a\sin\theta}{\varpi}$, both of which have multiplicity 2 and
are positive everywhere on this space with Zipoy topology.
	(Note that $\lambda_- \to 0$ on the ring, which is not part of the space time but at its boundary.)
	We may thus take the above as the definition of a positive definite inner product
 given by the matrix $\hat{M}$ for bispinors defined on any $t=const.$ section of $\cM$, a rectangular cylinder
${\mathcal{Z}} :=\Rset\times [0,\pi]\times  [0,2\pi]$  with its natural measure:
\begin{equation}\label{def:innerPROD}
\langle \hat{\Psi},\hat{\Phi}\rangle_{\hat{M}} := \int_{{\mathcal{Z}}} \hat{\Psi}^\dag\hat{M} \hat{\Phi} d\theta d\varphi dr.
\end{equation}
 The corresponding Hilbert space is denoted by ${\sf H}$, thus
\begin{equation}
{\sf H} 
:= \left\{ \hat\Psi:{\mathcal{Z}} \to \Cset^4\ | \ \|\hat\Psi\|_{\hat{M}}^2 := \langle \hat\Psi,\hat\Psi\rangle_{\hat{M}} < \infty \right\}.
\end{equation}
 Note  that ${\sf H}$ is \emph{not equivalent} 
to $L^2({\mathcal{Z}})$ whose inner product has the identity matrix in place of $\hat{M}$.

 We are finally ready to list our main results which are proved in \cite{KTZzGKNDa}.

\subsubsection{Symmetry of the Spectrum of $\hat{H}$}

Following the strategy of Glazman, in \cite{KTZzGKNDa} we prove:

\begin{thm}\label{thm:sym}
  Let any self-adjoint extension of the formal Dirac operator $\hat{H}$ on $\sf H$ be denoted by the same letter.
  Suppose $E\in\mathrm{spec}\,\hat{H}$. 
  Then $-E\in\mathrm{spec}\,\hat{H}$. 
\end{thm}

\begin{rem} We can also replace $\textsc{q}a$ with $\textsc{i}\pi a^2/c$ in \eqref{zGKNA}, introducing 
a KN-anomalous magnetic moment; here $\textsc{i}$ is an electrical current supported by the ring singularity, 
independently of $\textsc{q}$.
 This changes $\tilde{A}_0$ and $\tilde{A}_2$ to
\begin{equation}\label{generalKNelmagAtilde}
\tilde{A}_0=-\textsc{q}\frac{r}{|\rho|\varpi}-\left(\textsc{q}-{\textsc{i}\pi a/c}\right)\frac{a^2r\sin^2\theta}{\varpi |\rho|^3},\
\tilde{A}_2 = - \left(\textsc{q}-\textsc{i}\pi a/c\right)\frac{ar\sin\theta}{|\rho|^3}.
\end{equation}
Our symmetry result Thm. \ref{thm:sym} holds for \emph{any} self-adjoint extension of $\hat{H}$, whatever $\textsc{q}$ and 
$\textsc{i}$.
\hfill ${\qed}$
\end{rem}

\subsubsection{Essential Self-Adjointness of $\hat{H}$}

\hspace{-4pt}
 Adapting an argument of Winklmeier--Yamada \cite{WinYamB}, in \cite{KTZzGKNDa} we prove:
\begin{thm}\label{thm:esa}
Let $\textsc{q} = e = \textsc{i}\pi a/c$.
 Let ${\mathcal{Z}}^\circ$ denote ${\mathcal{Z}}$ with the ring singularity $\{(r,\theta,\varphi)|r=0,\theta=\pi/2\}$ deleted.
Then $\hat{H}$ with domain $C^\infty_c({\mathcal{Z}}^\circ)$ is essentially self-adjoint
on~$\sf H$.  
\end{thm}

 The unique self-adjoint extension of $\hat{H}$ will also be denoted by $\hat{H}$.

\subsubsection{The Continuous Spectrum of $\hat{H}$}

Using the Chandrasekhar--Page--Toop separation of variables, and an argument of Weidmann \cite{Wei82},
in \cite{KTZzGKNDa} we prove:

\begin{thm}\label{thm:essspec}
For $\textsc{q}= e=\textsc{i}\pi a/c$, the continuous spectrum of $\hat{H}$ on $\sf H$ is $\Rset\setminus(-mc^2,mc^2)$.
\end{thm}

\subsubsection{The Point Spectrum of $\hat{H}$}

In \cite{KTZzGKNDa} we prove:

\begin{thm}\label{thm:ptspec}
Let $\textsc{q} = e = \textsc{i}\pi a/c$.
If 
$2|a|<\frac{\hbar}{mc}$ and $\frac{e^2}{\hbar c} <\sqrt{\frac{2|a|}{\hbar/mc}\big(1-\frac{2|a|}{\hbar/mc}\big)}$,
the point spectrum of $\hat{H}$ on $\sf H$ is nonempty and located in $(-mc^2,mc^2)$; the end points are not included.
\end{thm}

\begin{rem}
 In the \emph{hydrogenic} problem where the ``proton'' charge $e$ is replaced by the
charge $\textsc{q}=\textsc{z}e$ of a ``nucleus,'' with $\textsc{z}>1$ (and the proton mass $m_{\mbox\tiny{p}}$ by
the nuclear mass $\textsc{m}>0$ --- although this does not show in the zero-$G$ formulation of the problem 
of a ``test'' electron in the electromagnetic field of a nucleus),
a point spectrum exists in the gap of the continuum as long as
$\textsc{z}< 137.036 \sqrt{\frac{2|a|}{\hbar/mc}\big(1-\frac{2|a|}{\hbar/mc}\big)}$.
 Since our upper bound on $\textsc{z}$ goes $\downarrow 0$ as $|a|\downarrow 0$, it is presumably 
not sharp, at least not when judged against the familiar Dirac bound $\textsc{z}<137.036$ for the existence of a point spectrum 
in the hydrogenic problem with point nuclei on Minkowski spacetime.
\hfill ${\qed}$
\end{rem}

We briefly indicate our strategy of proof of the point spectrum.
 We employ the Chandrasekhar--Page--Toop separation of variables Ansatz, 
\begin{equation}
\Psi(t,r,\theta,\varphi) = e^{-iEt+i\kappa \varphi} \left( \begin{array}{c}R_1(r)S_1(\theta)\\ 
     R_2 (r)S_2(\theta)\\ R_2(r)S_1(\theta)\\ R_1(r) S_2(\theta) \end{array}\right),
\end{equation}
with $E\in(-mc^2,mc^2)$ and $2|\kappa|\in\Zset\setminus\{0\}$,
obtaining coupled eigenvalue problems for $\vec{R}=(R_1,R_2)^t$ and $\vec{S}=(S_1,S_2)^t$,
\begin{equation}
T_{rad}\vec{R} =  E\vec{R},\qquad T_{ang}\vec{S} = \lambda \vec{S},
\end{equation}
where
\begin{equation}
T_{rad} = \left(\begin{array}{cc}  i \frac{d}{dr} - \frac{-a\kappa + eQ r}{\varpi^2} 
&-m\frac{r}{\varpi} - i\frac{\lambda}{\varpi} \\ -m\frac{r}{\varpi}+i\frac{\lambda}{\varpi} 
& - i \frac{d}{dr} -\frac{-a\kappa + eQ r}{\varpi^2} \end{array}\right)
\end{equation}
and
\begin{equation}
T_{ang}= 
\left(\begin{array}{cc}  -ma\cos\theta & -\frac{d}{d\theta} - \left( aE\sin\theta - \frac{\kappa}{ \sin\theta}\right)\\
\frac{d}{d\theta} - \left( aE\sin\theta - \frac{\kappa}{\sin\theta}\right) &ma\cos\theta  \end{array}\right).
\end{equation}

 The Pr\"ufer transform 
\begin{equation}
R_1 = Re^{i\Omega/2},\   R_2 = Re^{-i\Omega/2},\  S_1 = S \cos\frac{\Theta}{2},\ S_2 = S \sin\frac{\Theta}{2}
\end{equation} 
now yields a partly decoupled nonlinear eigenvalue system, 
\begin{equation}
\left\{\begin{array}{l}
\!\!
d\Omega/dr = 2 \frac{mr}{\varpi} \cos\Omega + 2\frac{\lambda}{\varpi} \sin\Omega +2\frac{a\kappa + \gamma r}{\varpi^2} - 2E 
\label{Oeq}
\\
\!\!
d(\ln R)/dr= \frac{mr}{\varpi}\sin\Omega - \frac{\lambda}{\varpi} \cos\Omega 
\end{array}\right.
\end{equation}
\begin{equation}
\left\{\begin{array}{l}
\!\!
d\Theta/d\theta = 2(\lambda -ma\cos\theta\cos\Theta + \left(aE \sin\theta - \frac{\kappa}{\sin\theta}\right)\sin\Theta) 
\label{Teq}
\\
\!\!
d(\ln S)/d\theta = -ma \cos\theta\sin\Theta - \left(aE\sin\theta - \frac{\kappa}{\sin\theta}\right)\cos\Theta.
\end{array}\right.
\end{equation}
 Note that in each pair of equations the second one can be integrated once a solution to the first one is known.
 The first equation in each pair is independent of the second one in the pair; however, the two first equations are still
coupled through the eigenvalue parameters and need to be solved jointly.

 There are integrability conditions.
 Combined with the Chandrasekhar et al. Ansatz the Pr\"ufer transform yields
\begin{equation}
\Psi(t,r,\theta,\varphi) = R(r)S(\theta)e^{-i(Et-\kappa \varphi)} \left(\begin{array}{l}
\cos(\Theta(\theta)/2)e^{+i\Omega(r)/2}\\
\sin(\Theta(\theta)/2) e^{-i\Omega(r)/2}\\
\cos(\Theta(\theta)/2)e^{-i\Omega(r)/2}\\
\sin(\Theta(\theta)/2)e^{+i\Omega(r)/2}\end{array}\right),
\end{equation}
and $\Psi \in L^2$ iff: 
\begin{equation}\label{OTasymp}\left.
\begin{array}{rllrlr}
& \Omega(-\infty) & =\quad -\pi + \cos^{-1}(E), &\Omega(\infty) &=\quad - \cos^{-1}(E)\\
& \Theta(0) &=\quad 0, & \Theta(\pi) &=\quad -\pi.
\end{array}
\right\}
\end{equation}
 
 The two coupled equations \eqref{Oeq}, \eqref{Teq}, supplemented by the asymptotic conditions \eqref{OTasymp}, can be
interpreted as a dynamical system and treated as such with dynamical systems theory; for the many details, 
see \cite{KTZzGKNDa}.

This completes the survey of our main results from \cite{KTZzGKNDa}.

\begin{figure}[ht]
\begin{center}\includegraphics[scale=0.3]{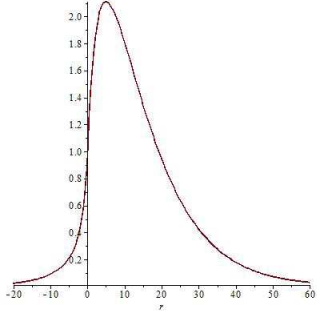}\end{center}
\caption{The absolute square of the (putative) positive ground state vs. the oblate spheroidal $r$ coordinate.
\label{fig:PSIg}}
\end{figure}

 We have also numerically computed (what we believe is) the positive energy ground state for various $a$ values; 
a typical profile is shown in Fig.~\ref{fig:PSIg}.

 For small $|a|$ the profile is close to the known Hydrogen ground state profile for $a=0$ in the $r>0$ sheet; 
in addition, a tiny exponentially decaying ``tail'' extends into the $r<0$ sheet. 
 This is what one would intuitively expect from a negatively charged electron in the $r>0$ sheet that is attracted to a small-$a$ 
ring singularity which appears positively charged in the $r>0$ sheet: the likelihood of finding the electron on the negatively
charged side of the ring singularity should be exponentially small. 
 Interestingly, by anti-symmetry, the negative energy ``ground'' state (in the sense of having smallest absolute energy) 
is obtained by reflection of the positive energy ground state profile at $r=0$.
 By the same kind of ``intuitive explanation'' given of the behavior of the electron in the positive energy ground
state, the negative energy ground state corresponds to the behavior expected rather from a positron. 

\section{Summary and Outlook}

\subsection{Summary}
 Motivated by the mathematical-physics problem of general-relativistic extensions of the Sommerfeld 
fine-structure spectrum of Hydrogen, in particular those including the hyper-fine structure,
in this presentation we have addressed the Dirac equation of a point electron in the zero-$G$ limit of the 
maximal analytically extended, double-sheeted Kerr--Newman spacetime.
 A related study has been proposed by Pekeris \cite{Pek87} who, however, studied Dirac's equation 
on a \emph{one-sheeted truncation} of the z$G$KN spacetime, which comes at the price of nonintegrable 
``proton'' charge and current ``densities'' concentrated in a disc\footnote{For integrable yet infinitely extended
astrophysical Kerr--Newman disc sources, see \cite{LedZofBic98}.}
 and raises the question of boundary
conditions for the Dirac bispinors at the disc; cf. also \cite{GairPRIZE}.
 By contrast, we have found that our z$G$KN Dirac Hamiltonian is essentially self-adjoint, and its unique self-adjoint
extension has a spectrum which is symmetric about zero, containing the familiar continuum $(-\infty,-mc^2)\cup(mc^2,\infty)$ 
plus, under a smallness condition, a discrete spectrum in the gap of the continuum;
for the Hydrogen parameter values the smallness condition is satisfied.
 Moreover, our results imply that the point spectrum converges to two anti-symmetric copies of the Sommerfeld spectrum when
the ring radius of the z$G$KN spacetime vanishes, if it converges to the spectrum of the zero-$a$ operator.
 
 We end this summary with a disclaimer: 
we are not advocating that the ring singularity of the double-sheeted z$G$KN spacetime were an accurate
model for a physical proton.
 Rather our z$G$KN Born-Oppenheimer Hydrogen Atom model is merely an interesting toy model which reduces to the 
familiar special-relativistic Born--Oppenheimer Hydrogen Atom model (and an antisymmetric copy thereof) when $a = 0$, 
and thus allows one to rigorously study non-perturbatively some general-relativistic $a>0$ effects on the quantum-mechanical 
Hydrogen spectrum, such as the hyperfine structure (which cannot be studied non-perturbatively with the point proton model,
featuring an electric charge and a magnetic dipole moment).
\vspace{-3pt}

\subsection{Outlook}

 Our study leaves many questions unanswered, but also suggests some intriguing speculations (the latter were not included in the
talk at Regensburg).
\vspace{-2pt}

\subsubsection{Open Questions}
 First of all, we would like to know the point spectrum of the ``double-sheeted z$G$KN'' Dirac Hamiltonian in more detail,
and as function of the ring radius $a$; a numerical study is currently in progress.

 More in line with PDE research into Dirac's equation on general relativistic spacetimes, we would like to know what
happens when a Dirac bispinor wave function impinges on the z$G$KN ring singularity; in particular, how much of it scatters and
how much will dive through the ring? 

 Furthermore, we would like to know what happens if the z$G$KN magnetic moment $\textsc{q}a$ is replaced by $\textsc{i}\pi a^2/c$,
so that the electromagnetic z$G$KN spacetime becomes a z$G$K spacetime decorated with an Appell--Sommerfeld field of arbitrary
charge $\textsc{q}$ and current $\textsc{i}$. 
 Is the Dirac Hamiltonian still essentially self-adjoint? 
 If not, are there distinguished self-adjoint extensions?
 Can one characterize the spectrum of the self-adjoint extension(s)? 
 Note that, if $\textsc{q}\neq \textsc{i}\pi a/c$ then the Chandrasekhar--Page--Toop Ansatz to separate variables fails, and 
one is faced with a two-variable PDE eigenvalue problem.

 Incidentally, independent of any inquiry into the Dirac equation, the following question is relevant to the problem 
of uniqueness of the Kerr--Newman manifold:
is it possible to $G$-deform the z$G$K spacetime decorated with an Appell--Sommerfeld field of 
charge $\textsc{q}$ and current $\textsc{i}$ into a solution of the Einstein--Maxwell equations only for the KN choice
$\textsc{q}a =\textsc{i}\pi a^2/c$?

 Thus we also would like to investigate what happens when gravity is ``switched back on.'' 
 After reviewing the enormous obstacles which are encountered in the pertinent literature on the subject, we concluded that
these are caused mainly by the non-integrable electromagnetic self-energy densities of point electron and point or ring proton as 
computed with the linear Maxwell--Lorentz equations which lead to unphysical spacetime curvatures once gravity is switched on. 
 To avoid these problems we plan to study the nonlinear Einstein--Maxwell--Born--Infeld system, which promises to yield the mildest
conceivable spacetime singularities \cite{TZeEMBI}. 
 Unfortunately, their nonlinearity is formidable, and progress will most likely come slowly. 
 Moreover, it is clear that one has to abandon the Born--Oppenheimer approximation, but 
the full two-body problem (cf. \cite{BeSaBOOK}) may still be out of reach.
\vspace{-4pt}

\subsubsection{Speculations}
 As for the intriguing speculations, our research also led to a completely different line of inquiry which we have embarked on in
\cite{KTZzGKNDb}.
 Namely, as has been advocated by St\"uckelberg \cite{Stueckelberg42}, 
Feynman \cite{Fey49}, Thaller \cite{Thaller}, \cite{ThallerBOOK}, and others, the puzzling spectral 
properties of Dirac's equation interpreted quantum mechanically, as again highlighted by the mysterious anti-symmetry of the
Dirac spectrum  of a ``point electron'' in the anti-symmetric double-sheeted z$G$KN spacetime, suggest that Dirac's equation 
captures the dynamics and bound states of both \emph{electron and positron}. 
 Yet interpreted as a quantum-mechanical equation,  Dirac's equation is  a \emph{single particle} equation. 
 This has suggested to us to entertain the hypothesis that particles and anti-particles are merely  ``different sides of 
the same medal,'' i.e. forming a single meta-particle with a binary structure, rather than being different particles in 
their own right.
 The very anti-symmetric structure of the z$G$KN ring singularity supplies just such a binary structure, which in \cite{KTZzGKNDb}
we have tentatively identified with an electron / anti-electron meta-particle.
 There we show that the Dirac spectrum of such a z$G$KN-ring particle in the electrostatic field of a 
given point charge (now playing the role of the point proton) with straight world line in the pertinent 
z$G$KN spacetime is determined by the same equation that we have discussed in this presentation. 
 It's the narrative that changes, not the mathematics.
 This narrative, where electron and anti-electron are just two different ``sides of the same medal,'' is
faithfully realized by the electromagnetic ring singularity of the z$G$KN spacetime, and we ponder seriously
the possibility of it having a true physical significance.


\subsection*{Acknowledgment}
Many thanks go to Felix Finster, J\"urgen Tolksdorf, and Eberhard Zeidler for their kind invitation to present
these results at their superbly organized conference, and for the financial support and the impeccable hospitality 
offered by the organizers and their staff.  We also thank Donald Lynden-Bell and Jonathan Gair for the permission to 
reproduce their field line drawings (Fig.~\ref{fig:KNfieldsEandB}). Finally, we thank the referee for a very careful 
reading of our paper, and for constructive criticisms.

\end{document}